\documentclass[sigconf]{acmart}
\usepackage{xcolor}
\usepackage{graphicx}
\usepackage{array}
\usepackage{subfig}
\usepackage{makecell}
\usepackage{multirow}
\usepackage{amsmath}
\usepackage{mathtools}
\usepackage[justification=centering]{caption}
\AtBeginDocument{%
  \providecommand\BibTeX{{%
    \normalfont B\kern-0.5em{\scshape i\kern-0.25em b}\kern-0.8em\TeX}}}

\copyrightyear{2024}
\acmYear{2024}
\setcopyright{acmlicensed}\acmConference[Websci '24]{ACM Web Science Conference}{May 21--24, 2024}{Stuttgart, Germany}
\acmBooktitle{ACM Web Science Conference (Websci '24), May 21--24, 2024, Stuttgart, Germany}
\acmDOI{10.1145/3614419.3644009}
\acmISBN{979-8-4007-0334-8/24/05}

\begin{document}

\title{Inside the echo chamber: Linguistic underpinnings of misinformation on Twitter}


\author{Xinyu Wang}
\affiliation{%
  \institution{The Pennsylvania State University}
  \city{University Park, PA}
  \country{USA}}
\email{xzw5184@psu.edu}

\author{Jiayi Li}
\affiliation{%
  \institution{The Pennsylvania State University}
  \city{University Park, PA}
  \country{USA}}
\email{jpl6207@psu.edu}

\author{Sarah Rajtmajer}
\affiliation{%
  \institution{The Pennsylvania State University}
  \city{University Park, PA}
  \country{USA}}
\email{smr48@psu.edu}

\renewcommand{\shortauthors}{Wang, et al.}

\begin{abstract}
Social media users drive the spread of misinformation online by sharing posts that include erroneous information or commenting on controversial topics with unsubstantiated arguments often in earnest. Work on echo chambers has suggested that users' perspectives are reinforced through repeated interactions with like-minded peers, promoted by homophily and bias in information diffusion. 
Building on long-standing interest in the social bases of language and linguistic underpinnings of social behavior, this work explores how conversations around misinformation are mediated through language use. 
We compare a number of linguistic measures, e.g., in-/out-group cues, readability, and discourse connectives, within and across topics of conversation and user communities. 
Our findings reveal increased presence of group identity signals and processing fluency within echo chambers during discussions of misinformation.  We discuss the specific character of these broader trends across topics and examine contextual influences.
\end{abstract}






\maketitle
\section{Introduction}
In recent years, Online Social Networks (OSNs) have developed into a primary source of news for a growing majority of the US population \cite{levy2021social}. As OSNs have expanded, so have concerns about the role of social media in promoting misinformation and deepening social fissures \cite{allcott2019trends,zimdars2020fake}, the consequences of which were particularly tangible during the COVID-19 pandemic and response \cite{abrams2020mitigating,brennen2020types,bursztyn2020misinformation}.

User interactions in OSNs are scaffolded by platforms' algorithms and moderation practices. 
Optimized primarily for engagement, these algorithms promote interaction with like-minded peers and isolate users from disagreeable content. While this can create safe space for individuals to share opinions and find validation, it can also drive polarization \cite{barbera2020social,cinelli2021echo,rhodes2022filter}. Some have suggested that the effects of so-called \emph{filter bubbles} and \emph{echo chambers} have been overstated or oversimplified and have argued for a more nuanced characterization based on, e.g., ideology or topic of discussion \cite{barbera2015tweeting,bruns2017echo,guess2018avoiding}. Irrespective, these phenomena speak to the broader, well-studied phenomenon of homophily in social group formation \cite{mcpherson2001birds,currarini2009economic}. In the online setting, user-generated content facilitates aggregation around shared interests and worldviews, which in turn serves to support the 
deepening and reinforcement of social ties \cite{del2016spreading}.

At the heart of OSN communities is discourse rich with linguistic signals of individual and relational identities. That is, users of language perform their identities within the uses of language \cite{benwell2006discourse,zappavigna2014enacting,zhang2017community}. This is both heightened and complicated by the affordances of social media \cite{tagg2014language}. 
Identities are partially performed through alignment with communities, opinions, and cultural issues \cite{tagg2014language}. Online, this can be enacted as direct engagement and participation in groups, or through indirect interactions, e.g., hashtagging, sharing memes \cite{zappavigna2011ambient,zappavigna2014enacting}, or even simple repetition \cite{rosenblum20191}. 
%maybe here?

Several schools of thought have driven work to understand the co-evolution of language, social ties, and the environment \cite{five2009language,lupyan2010language,roberts2012social,bentz2018evolution,raviv2020role}. Languages adapt to the constraints and biases of their learners, which include social, physical and technological context \cite{dale2012understanding,lupyan2016there}. Foundational work in sociolinguistics, for example, has studied the ways in which language change correlates with social network dynamics \cite{milroy1992social,blommaert2010sociolinguistics}. Shared linguistic features can enhance so-called \emph{processing fluency}, making communications more effective and reinforcing community bonds \cite{canestrino2022impact,pancer2019readability}. Within the landscape of social media, both notions of community and language practices are flexible and interactively constructed \cite{tagg2014language}. 

Our work is a study of the relationships between social media users' linguistic patterns and the community structures in which they are embedded.
We have selected misinformation as our point of inquiry because prior work has suggested that misinformation spreads particularly quickly \cite{karlova2013social,kumar2014detecting,del2016spreading,shin2018diffusion,vosoughi2018spread}
and because a shared vocabulary has emerged around these topics.
We are motivated by two primary research questions.

\noindent\textbf{RQ1:}
Is there an observable increase in users' signaling of group identity through publicly shared content within echo chambers? Is this phenomenon more pronounced during discussions that involve topics related to misinformation?

\noindent\textbf{RQ2:} 
Is there heightened processing fluency in publicly shared content within echo chambers? Is this phenomenon more pronounced during discussions that involve topics related to misinformation?

To address these questions, we identify echo chambers from user interaction networks and study their linguistic characteristics. Our analysis suggests that while there are some shared linguistic patterns among certain topics, linguistic characteristics observed within echo chambers do not exhibit uniformity across all discussions of misinformation. Rather, these patterns appear to be contingent upon contextual factors (e.g., political nature of the topic). This interplay between linguistic uniformity and contextual particularity constitutes is at the core of our work, contributing to deeper understanding of the linguistic underpinnings of echo chambers, particularly in the context of misinformation-related topics.

\section{Related Work}

\subsection{Misinformation dynamics}
The extensive presence of user-generated content on online social media platforms enables the formation of communities that unite around shared beliefs and narratives. 
Echo chambers, where beliefs are amplified and reinforced by repeated internal validation, play a role in the dynamics of misinformation within online platforms \cite{del2016spreading}. The architecture of modern social media algorithms, which prioritizes user engagement, can inadvertently foster echo-chambered communities \cite{zollo2015emotional}.
Within these environments, misinformation thrives and evolves, taking on new narratives that further reinforce the community's existing beliefs \cite{vosoughi2018spread}. 

\subsection{Social identity formation}
Social identity formation emerges as a critical process underlying group dynamics and individual affiliations in social networks. In this study, social identities are performed and measured through a combination of explicit identity cues, as explained by social identity theory, and implicit identity cues via ambient affiliation.

\subsubsection{Social identity theory.}
Social identity theory posits that individuals tend to see themselves and others in relation to groups \cite{tajfel1979integrative,tajfel2004social,stets2000identity}. When a group identity becomes prominent, individuals develop their self-concept around paradigmatic ingroup characteristics \cite{brewer1999psychology}. They are motivated to differentiate between their ingroup and relevant outgroups \cite{li2020real}. Social identity theory has been engaged to describe political group formation in online social communities, e.g., \cite{greene2004social,trepte2017social}. 

Recent work \cite{li2020real} looked at discussions containing the term ``fake news'' on Twitter and found an uptrend in the use of identity language, as measured by increased frequency of group pronouns. They also found that group pronouns were likely to collocate with words boosting ingroup messaging. These findings dovetail with concerns about polarization. Rathje et al. \cite{rathje2021out} found out-group animosity to be a predictor of engagement on Facebook and Twitter, due in part to algorithmic designs favoring messages containing out-group antagonism. Our work operationalizes linguistic indicators of social identity as in-/out-group reference language.

\subsubsection{Ambient affiliation.}
So-called \emph{ambient affiliation} through hashtagging has evolved as a ubiquitous mechanism for social media users to associate and communicate in digital environments, especially on OSNs. It allows for the spontaneous aggregation of like-minded individuals around shared interests, experiences, or sentiments, often leading to the establishment of virtual communities or ``micro-groupings" \cite{zappavigna2011ambient}. More than just categorization, the act of hashtagging serves as an instrument of social group identity formation. By selecting specific hashtags, users signal an allegiance or a stance, fostering both inclusion within communities and distinction from others. Consistent patterns of hashtagging behaviors can solidify these group identities, creating online communities unified by shared narratives and ideologies \cite{zappavigna2012discourse,zappavigna2014enacting}.

\subsection{Processing fluency}
Processing fluency theory suggests that the subjective experience of ease with which people process information influences people’s judgments of that information \cite{reber1998effects,reber2012processing}. People are more likely to judge information favorably if they are able to digest it easily. This effect has been shown robust over diverse experimental settings and a variety of formulations of fluency, e.g., conceptual, perceptual, linguistic \cite{oppenheimer2008secret,alter2009uniting}. Central to this theory is the notion that enhanced fluency can be interpreted in multiple ways: it might signify the benefits of a simplified message; the role of priming and familiarity; or, the importance of explanation and elaboration.

\subsubsection{Simplicity bolsters information processing.}
 
Recent work has shown that easy-to-read social media posts facilitate processing fluency, resulting in more likes, comments, and shares \cite{bailey2013increasing,pancer2019readability}. 
Likewise, repeated assertions gain a valence of credibility \cite{hasher1977frequency,reber2010epistemic,unkelbach2017referential}. The so-called repetition-induced truth effect explains one mechanism by which individuals come to trust obviously false information, a concern magnified by echo chambers \cite{pennycook2018prior,unkelbach2021mere}. In the context of today's misinformation landscape, scholars have noted that conspiracism is increasingly expressed as bare assertion \cite{rosenblum20191}. New conspiracists repeat simple statements, signaling solidarity with one another through retweets, likes, and shares \cite{rosenblum20191}.

\subsubsection{Elaboration bolsters information processing.}
In parallel, there exists a body of psychological theories suggesting that explanations bolster credibility of a claim, namely, the idea of \emph{rational persuasion}. Information-processing theories of human cognition have explored the ways in which we engage in the reception, encoding, storage, and retrieval of information \cite{simon1978information,anderson1980cognitive}. Systematic information processing is associated with explanations, whereby individuals engage in effortful scrutiny and comparison of information \cite{trumbo1999heuristic,todorov2002heuristic,chaiken2012theory}.
Relatedly, the Elaboration Likelihood Model (ELM) \cite{petty1986communication,petty1999elaboration} suggests that when people process information deeply, they are more likely to be influenced by it. 
Prior studies on the QAnon conspiracy have demonstrated the efficacy of this technique. Individuals aligned with these beliefs use archival practices to enable the construction and diffusion of knowledge within their community. Members are urged to engage in critical thinking and do extensive research using key terms promoted by the group \cite{marwick2022constructing}.

\section{Data}

We collected tweets using the Twitter Developer API and keywords associated with our target topics, namely: anti-vaccination; 2020 US election fraud; and, QAnon. We pulled author ID, tweet ID, and referenced tweets (type and ID) for interaction network construction and tweet content for linguistic analyses. 
Tweets were collected using popular keywords and hashtags over date ranges selected for greatest topic relevance. %that the topic was mostly discussed. 
With the exception of discussion surrounding the US election which has a shorter activity period and limited number of tweets, we collect approximately 1 million tweets per topic, adjusting date ranges accordingly.  % to avoid huge discrepancy in community size.  
In addition, we collected discussions to serve as comparative for our misinformation-related discourse: the MeToo movement, Black Lives Matter movement (BLM), and \#mondaymotivation. While not generally associated with misinformation, these have been topics of widespread social media discourse, representing very different degrees of social sensitivity and polarization.

\noindent \emph{Pre-processing.} We retained only tweets containing interactions within the datasets, as well as original tweets without interactions, to ensure that all users in the social network have representative linguistic content.
Dataset statistics are provided in Table~\ref{tab:table_stats}. 
We removed \@mentions, \#hashtags, URLs, RT signs, and emojis. For readability tests, we retained punctuation marks to allow sentence length calculation. Hashtags and URLs were extracted directly from the raw data. Tweet IDs and code have been made publicly available and are provided with detailed documentation including a Datasheet \cite{gebru2021datasheets} and README\footnote{Github link: https://github.com/XinyuWang1998/Linguistic-Underpinning}.

\begin{table*}[ht!]
    \centering
    \small
    \begin{tabular}{|c|c|c|c|c|c|}
        \hline
        \textbf{Category}&\textbf{Topic} &\textbf{keywords (case-insensitive)}& \textbf{Time Frame} & \textbf{\makecell{Total\\ Tweets}} & \textbf{\makecell{Total Tweets\\ after Filtering}} \\
         \hline
          \multirow{7}{*}{Misinformation}&\makecell{US Election\\ Fraud}&\makecell{voterfraud, discardedballots,\\ cheatingdemocrats, stopvoterfraud,\\ voterfraudbymail, voterfraudisreal,\\ ballotharvasting, ballotvoterfraud}&\makecell{[2020-11-01,\\2021-02-28]}&256,036&219,872\\
         \cline{2-6}
          &\makecell{Anti-\\Vaccination}
          &\makecell{antivax, antivaxx, novaccine,\\ novaccinemandates, pureblood, unvaxxed,\\ unvaccinated, naturalimmunity, \\nomandatoryvaccines}&\makecell{[2021-11-15,\\2021-11-30]}&1,224,306&1,139,378\\
         \cline{2-6}
          
          &QAnon&\makecell{qanon, thegreatawakening, pizzagate, savethechildren,\\ deepstate, thestormisuponus}&\makecell{[2019-11-01,\\ 2020-03-31]}&902,862&792,733\\
         \hline\hline
          \multirow{8}{*}{\makecell{Hyper-polarized}}&\makecell{MeToo\\ Movement}&\makecell{metoo, hertoo, ustoo, metoocongress,\\ timesup \cite{xiong2019hashtag}}&\makecell{[2017-10-15,\\ 2017-10-31]}&845,472&828,856\\
          \cline{2-6}
          ~&\makecell{BLM \\Movement}&\makecell{blacklivesmatter,georgefloyd, icantbreathe,\\ blm, justiceforfloyd, justiceforgeorgefloyd,\\ georgefloydprotests, worldagainstracism,\\ walkwithus, kneelwithus, blackouttuesday,\\
voteouthate, nojusticenopeace,\\ blackwomenmatter,
blackgirlsmatter,\\ theshowmustbepaused\cite{field2022analysis}}&\makecell{[2020-06-01\\ 09:00AM,\\ 2020-06-01\\ 05:00PM]}&994,663&968,270\\
         \cline{1-6}
            \makecell{Neutral}&\makecell{Monday\\ Motivation}&monday motivation&\makecell{[2021-09-01,\\2021-11-30]}&896,350&892,096\\
         \hline
    \end{tabular}
    \normalsize
    \caption{Dataset statistics}
    \label{tab:table_stats}
\end{table*}

\section{Preliminaries}\label{sec:preliminaries}
We introduce key definitions and measures used throughout.

\subsection{Linguistic metrics}
Linguistic features were computed using and extending state-of-the-art open-source toolboxes for linguistic analyses. In- and out-group language was used to measure group identity formation. Group identity is expressed and sustained through pronouns; pronominal choices are not just categorical references but also reflect power dynamics and relationships \cite{inigo2004use, pennycook1994politics}. Specifically, ``we-pronouns" are used to invoke ingroup solidarity while ``they-pronouns" create distance with the outgroup, often implying the superiority of one's own community \cite{brewer1999psychology,li2020real}. We operationalize simplicity and elaboration as theoretical constructs through discourse connectives, big words, and readability metrics.
Research has shown that textually less complex and highly readable content facilitates processing fluency, resulting in more social interactive behaviors \cite{pancer2019readability}. While, discourse connectives may also improve online processing and reading comprehension in both narrative and expository texts \cite{zufferey2015advanced,crible2021lexical}. Evidence from these studies forms the basis for our approach.

\subsubsection{In-/out-group identity language.} In-/out-group language ratio \cite{crossley2016tool} is used as an index of group identity. Exemplars of in-/out-group cues include: we, our, us, they, their, them, themselves. 

\subsubsection{Big words.} 
Words containing seven or more letters serve as a rough indicator of language complexity. 
This metric is derived from The Linguistic Inquiry and Word Count (LIWC) \cite{LIWC} software. 

\subsubsection{Readability.}
We employ the Flesch reading ease score \cite{flesch1948new} as a measure of readability based on two factors: \emph{average sentence length} and \emph{average number of syllables in each word}. 

\subsubsection{Discourse connectives.} We consider the incidence of discourse connectives \cite{addawood2019linguistic,crossley2016tool} as an indicator of explanation/elaboration. Our analyses consider three families of discourse connectives. 

\noindent\textbf{Logical connectives} reflect logical explanation, e.g., because, consequently, hence; 

\noindent\textbf{Additive connectives} reflect agreement and further explanation, e.g., and, besides, further; 

\noindent\textbf{Negative connectives}, indicate an adversative argument, e.g., but, alternatively, although.

\noindent Connectives ratios are calculated as the number of connective words from the given family divided by total number of words. 

\subsection{Network representation}
We construct user interaction networks for each topic. 
Users are represented as nodes and interactions--replies, quotes, retweets--are represented as directed, weighted edges. Edge weights indicate number of interactions.

\textbf{Echo chambers} are recognized by coherent content and structural connectivity. Given that our interaction networks are topic-specific by virtue of our data collection, there is coherence in content by design. For structural identification of echo chamber members, we focus on \emph{strongly connected components}, i.e., subgraphs in which there is a path from each vertex to every other, respecting directionality of the edges. This approach follows precedent set in prior literature, e.g., \cite{cota2019quantifying}, and is particularly suitable for modeling echo chambers on platforms like Twitter where interactions are often non-reciprocal \cite{kwak2010twitter,smith2012does}. 
Following, we identify users within strongly connected components ($n\geq2$) as echo chamber members and their shared content as echo chamber tweets. User statistics are provided in Table~\ref{tab:EC_stats}.

\subsection{Ambient affiliation}
Sociolinguists have theorized that Twitter users signal group identity implicitly through hash-tagging. Hashtags serve as a semiotic identifier for the discourse community they establish, as well as a searchable record of discourse for the purpose of establishing new affiliates \cite{zappavigna2011ambient}. We measure \textbf{ambient affiliation} through the use of hash-tagging.

\begin{table}[!ht]
    \centering
    \begin{tabular}{|c|c|c|c|}
    \hline
        \textbf{Topic} & \textbf{EC} &\textbf{\makecell{EXT}} & \textbf{\makecell{Total}}\\ \hline
        Election Fraud&470&97,510&97,980 \\ \cline{1-4}
        Antivax&18,200&393,331&411,531\\ \cline{1-4}
        QAnon&8,934&233,166&242,100 \\ \hline\hline
        MeToo&3,468&449,627&453,095  \\ \cline{1-4}
        BLM&1,505&664,930&666,435  \\ \cline{1-4}
        Monday Motivation &3,019 & 339,903&342,922\\ \hline
    \end{tabular}
    \normalsize
    \caption{Number of users/nodes engaged with each topic (EC=echo chamber; EXT=external).}
    \vspace{-0.5cm}    
    \label{tab:EC_stats}
\end{table}

\section{RQ1: Group identity formation}
RQ1 asks whether there are observable disparities in group identity between echo chambers and external groups within discussions surrounding misinformation-related topics. We assess this through two approaches: analyzing explicit signaling of group identity through the use of first and third-person pronouns; and, exploring implicit signaling of group identity through ambient affiliation.

\subsection{Explicit identity signaling through in-/out-group cues}

\subsubsection{Methodology.}

We employ the zero-inflated beta regression model to examine the relationship between a user's membership to (presence within) an echo chamber and their use of group pronouns. The zero-inflated beta model is particularly suited for this analysis due to the nature of the data, which contains many zeros and is bounded between 0 and 1. We concentrate on the model's non-zero component because it remains largely unaffected by extreme outliers, specifically instances where the content is very short thus not containing target vocabularies. The coefficient of the predictor variable in this regression, which represents echo chamber membership, provides an estimate of the log odds ratio of the change in identity measurement in response to echo chamber membership. If this coefficient is positive and statistically significant, it suggests that membership to an echo chamber is associated with higher log odds of using group pronouns. Recognizing that analyses with large datasets often yields small, less informative p-values, we employ a stratified random sampling method \cite{fox2002bootstrapping}.  This involves selecting 1000 data points each from echo chamber and non-echo chamber data subsets (2000 total), ensuring a balanced representation of both categories. To address potential small sample biases and enhance the robustness of our findings, we conduct 1,000 iterations of this regression analysis, obtaining median values for the coefficient, standard error, T-score, and p-value. Analyses without sampling are also provided, for completeness, alongside datasets and code through the project's Github page.

\subsubsection{Findings.}
\textbf{We observe clear, heightened group identity development through the deliberate utilization of target plural pronouns in echo chambers within US election fraud and QAnon discussions.} See Table~\ref{tab:identity}. 

Following is a typical tweet from within a QAnon echo chamber that exemplifies pronounced group identity. After mentioning 50 names in a chain, the tweet reads:
\begin{quote}
\textit{Totally agreed. We Americans chart our own destiny
The Individual is THE most protected of its rights by the Constitution
we need to know this, and go forward confidently.
DeepState, we will deal w/them in unity w/our President.
For such a time like this-we have our fighting man!}
\end{quote}

\noindent Likewise, strong group identity cues are evident within election fraud echo chambers. For instance:
\begin{quote}

      \textit{@realDonaldTrump keep fighting like the warrior you are! We are with you! \#VoterFraud https://t.co/RuxhsXyO8U}   

\end{quote}

Note that we observe an inverse coefficient for Monday Motivation. We expect this may be attributed to the inherent characteristics of the challenge, wherein individuals share personal experiences with their communities \cite{ridings2002some}. 

\subsection{Implicit identity signaling through ambient affiliation}

\begin{figure*}[ht]
    \centering
    \includegraphics[width=\linewidth]{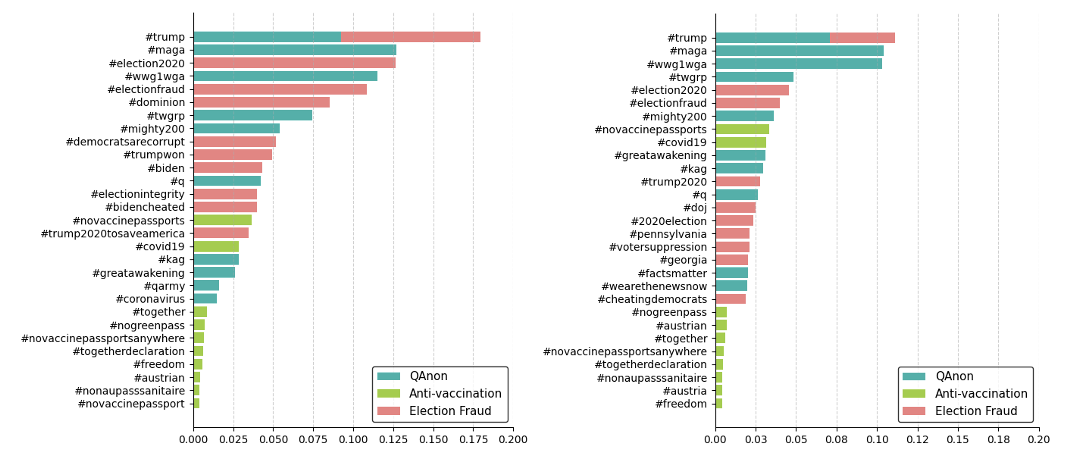}
    \caption{Top 10 hashtags, by density, in echo chambers (left) and overall (right) for misinformation topics.}
    \label{fig:top_hashtags}
\end{figure*}

\begin{table*}[!ht]
    \centering
    \begin{tabular}{|c|c|c|c|c|c|}
    \hline
        \textbf{Category} & \textbf{Topic} &\textbf{Coefficient} & \textbf{Std. Error}&\textbf{\makecell{T-score}}&\textbf{\makecell{p-value}}\\ \hline
        \multirow{3}{*}{Misinformation} & \textbf{Election Fraud} &\textbf{0.162}&\textbf{ 0.049}&\textbf{ 3.272}&\textbf{0.001**} \\ \cline{2-6}
        &Antivax &0.010&0.038&0.266&0.523  \\ \cline{2-6}
        ~ & \textbf{QAnon}&\textbf{0.116}&\textbf{0.045}&\textbf{2.339}&\textbf{0.019*} \\ \hline\hline
        \multirow{2}{*}{Hyper-polarized}&MeToo &0.026&0.044&0.593&0.453  \\ \cline{2-6}
        ~&BLM &-0.058&0.046&-1.265&0.206  \\ \cline{1-6}
        \makecell{Neutral} & \textbf{Monday Motivation} &\textbf{-0.186}& \textbf{0.056}&\textbf{-3.307}&\textbf{0.001**}\\ \hline
    \end{tabular}
    \normalsize
    \caption{Pairwise zero-inflated beta model median statistics for group identity language using bootstrap sampling. Positive coefficient represents positive association between echo chamber membership and increased group identity language.}
    
    \label{tab:identity}
\end{table*}

\subsubsection{Methodology.} 

Building on our observations of increased explicit group identity formation within echo chambers for misinformation-related topics in particular, we explore differences in ambient affiliation between echo chambers and the remainder of interaction network during discussions of misinformation-related topics. %This focused analysis is driven by the distinct tendencies in group identity expression with misinformation-related discussions, guiding us to investigate whether these patterns persist in the ambient affiliative behaviors.

\noindent\emph{Top hashtags.} 
We extract all hashtags present inside the dataset. 
From these, we select the 10 with highest density, standardized over the total number of tweets within each category. We exclude hashtags identical to the keywords used for data collection.

\noindent\emph{Hashtag networks.} We construct undirected common hashtag networks to model ambient affiliation within echo chambers, for all three topics of misinformation. In these networks, nodes represent users within echo chambers and a weighted edge exists between users who share at least one common hashtag. As before, we exclude hashtags identical to keywords utilized during data collection.

\subsubsection{Findings.} 

A conspicuous pattern emerges, complementing our findings from in-/out-group identity analyses. Both QAnon and election fraud discussions are characterized by methodical use of ambient affiliation relative to anti-vaccination dataset.  When conducting a comparison of hashtag usage within echo chambers versus overall hashtag usage, we observe that \textbf{echo chambers have notably higher rate of ambient affiliation.} This suggests members may be rallying around common hashtags to amplify their message and consolidate their identity. 

\vspace{0.1cm}
\noindent\emph{Top hashtags.}
It is clear that the strategic use of hashtags plays an important role, both within and outside echo chambers, in QAnon-related conversations (see Figure \ref{fig:top_hashtags}). Several hashtags speak directly to group membership, such as \#Mighty200 (a group of conservative or right-wing influencers) and \#wearethenewsnow (QAnon supporters use it to suggest that mainstream media cannot be trusted, positioning the QAnon community as the authentic purveyors of news). Some hashtags are slogan-like, e.g., \#WWG1WGA (``Where We Go One, We Go All") and \#kag (``Keep America Great").

\noindent\emph{Hashtag networks.}
Hashtag network statistics are provided in Table \ref{tab: hashtag_network}. In the QAnon echo chamber, the hashtag network exhibits the highest density, suggesting a deliberate and strategic approach to hashtag bulk use, likely for content promotion. Additionally, we see high percentage of ambient affiliation engagement and average common hashtags per paired users for election fraud, which is consistent with the high identity consolidation observed within election fraud discussions. Conversely, we observe limited hashtag usage within anti-vaccination discourse, which is also mirrored in the network structure. These findings highlight the complex nature of online discourse and nuance under the very broad umbrella of misinformation-related conversations.

\subsection{Case Study 1. Echo chambers in conversations about US election fraud}
We observe strongest group identity formation within US election fraud discourse. To further explore this, we conduct a case study analyzing interactions within, between, and outside echo chambers. %Figure~\ref{fig:fraud_ecnec} visualizes the interaction network between echo chamber members and external members. 
We find that echo chamber members are deeply embedded in the broader interaction network, maintaining active engagements with individuals outside their group. This heightened activity of echo chamber members in external interactions challenges the simplistic view of echo chambers as entirely insular entities and suggests a more nuanced understanding of discourse dynamics. 

\begin{figure}[ht]
    \centering
    \includegraphics[width=0.7\linewidth]{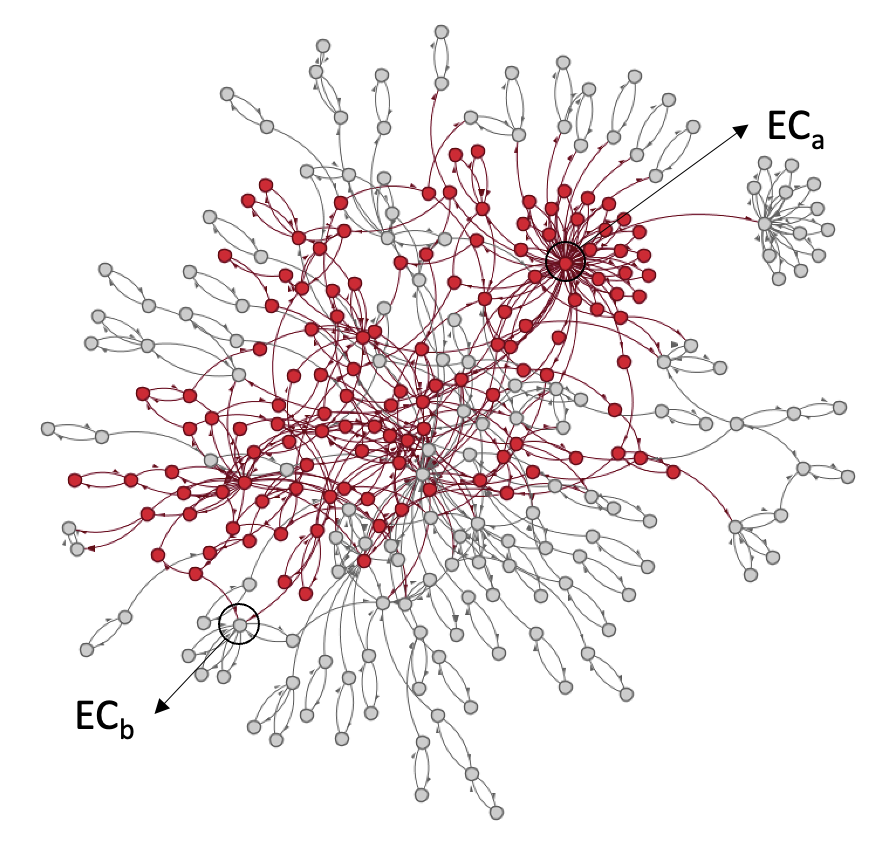}
    \caption{US election fraud. Internal interaction network amongst members of the largest echo chamber (red) and other echo chamber members with which they interact.}
    \label{fig:fraud_ec}
\end{figure}

%Next, we delve into the interactions taking place within the echo chamber community.
Figure~\ref{fig:fraud_ec} shows the internal interaction network amongst members of the largest echo chamber (red) and other echo chamber members with which they interact. We observe echo chambers are interconnected rather than isolated. %Among the 470 echo chamber nodes, 279 are connected through interactions. Within this subset of 279 connected nodes, 126 belong to the largest strongly connected component. 
Two nodes, in particular, exhibit highest centrality: Node EC\textsubscript{a}, a member of the largest echo chamber; and, EC\textsubscript{b}, a bridge between echo chambers.
Deeper examination of the content shared by these two nodes reveals distinct roles.
Node EC\textsubscript{a} emerges as a pivotal resource hub, disseminating information about the election status, often accompanied by a URL. These URLs typically redirect to tweets conveying similar content, supplemented by relevant hashtags.

Furthermore, we notice several indicators of group identity from EC\textsubscript{a} that extend beyond our initial examination of first and third-person plural pronouns. This content manifests as expressions of favoritism and support within the group or, conversely, as extreme animosity towards those outside the group. For instance:
 \begin{quote}
    \textit{Good morning @realDonaldTrump. I’m still here, standing with you and 70 million other freedom loving patriots. It time to expose the \#voterfraud. \#DoNotConcede \#DoNotConcedeUnderAnyCircumstances 
    \#DoNotGiveUp}

    \textit{GDMRNNG (The vowels have been stolen by Democrats) \#Election2020 \#VoterFraud \#BidenHarris2020 \#2020Elections} (Explanation: GDMRNNG is goodmorning without vowels.) 
      
\end{quote}

\noindent We observe that EC\textsubscript{b} mainly interacts with itself. Among 111 tweets posted by EC\textsubscript{b}, 93 are retweets or replies to its own tweets. Within this set of tweets, we notice: heavy use of user mentioning, an average of 46.32 mentions per tweet; use of hashtags, an average of 5.7 hashtags per tweet; and ending the post with a URL.

\begin{table*}[!ht]
    \centering
    \begin{tabular}{cccccc}
    \hline
        \textbf{Topic} &\textbf{\makecell{hashtag Network Density}}&\textbf{\makecell{Ambient Affiliation Engagement}}&\textbf{\makecell{Average Common\\ Hashtags per\\ Paired Users}}\\ \hline
        \makecell{Election Fraud}& 0.0721&52.34\% (246/470)&1.2327\\ \hline
        Anti-vax & 0.0716 & 8.37\% (1523/18200)&	1.0310  \\ \hline
        QAnon &0.0949&33.15\% (2962/8934)& 1.2123\\ \hline
    \end{tabular}
    \caption{Hashtag network statistics for three misinformation topics.}
    \label{tab: hashtag_network}
\end{table*}

\section{RQ2: Processing fluency effects}

We explore the impact of processing fluency on discourse around misinformation-related topics, with particular focus on the variance between echo chambers and external members. This study is not intended to compare the overall simplicity or elaboration between misinformation and control topics, but to identify differences in processing fluency between discussions within echo chambers and those outside of them. We utilize a series of linguistic metrics to measure processing fluency, complemented by an analysis of URLs, to understand these dynamics in detail.

\subsection{Processing fluency through linguistic simplicity}

\subsubsection{Methodology}

Simplicity is captured through readability and word length measurements.

\vspace{0.1cm}
\noindent \emph{Readability.}
As discussed in Preliminaries, we utilize the Flesch readability ease score. It considers not just word-level but also sentence-level complexity. Text processing follows as above, retaining all punctuation marks.
Calculation of the Flesch score requires text at least 100 words, so our approach deviated from standard data-point level comparison as follows. 
To address potential sampling bias, we employed a stratified sampling method wherein we selected 100 posts each from both the echo chamber and external post datasets and aggregated them for readability score calculation. This was repeated for 100 iterations. Given that the datasets exhibited a near-normal distribution, as depicted in Figure~\ref{fig:box_plot}, we utilize the T-test for analyzing mean differences. However, we observe non-uniform variance in certain instances, such as with the BLM topic. To reinforce the validity of our comparisons under these conditions, we additionally applied the non-parametric Mann-Whitney U test. Notably, the results from both testing methods converged.

\vspace{0.1cm}
\noindent\emph{Big words.}
We employ LIWC to calculate the ``bigword" score for each tweet. To avoid biases, we apply this analysis to cleaned texts, excluding hashtags, mentions, URLs, and similar elements that could be erroneously identified as ``big words."
Given that resulting scores are no longer bounded between 0 and 1, the zero-inflated beta model proves unsuitable for our comparative tasks. As the datasets being compared exhibit approximately normal distribution and similar variances, we employ an independent two-sample T-test. Given our relatively large dataset, this test tends to yield very small p-values, hence less informative. To mitigate this effect or potential non-normality in distribution, we perform bootstrapping by randomly sampling 1000 data points from both the echo chamber and non-echo chamber groups and conducting the t-test over 1000 iterations \cite{lumley2002importance,ahad2012relative,faber2014sample}. We select the median as the representative outcome.

\subsubsection{Findings}
\textbf{Our findings demonstrate clearly and consistently that simplicity, as a measure of processing fluency, is amplified within echo chambers in misinformation-related discussions.}

\begin{figure}[ht]
    \centering
    \includegraphics[width=\linewidth]{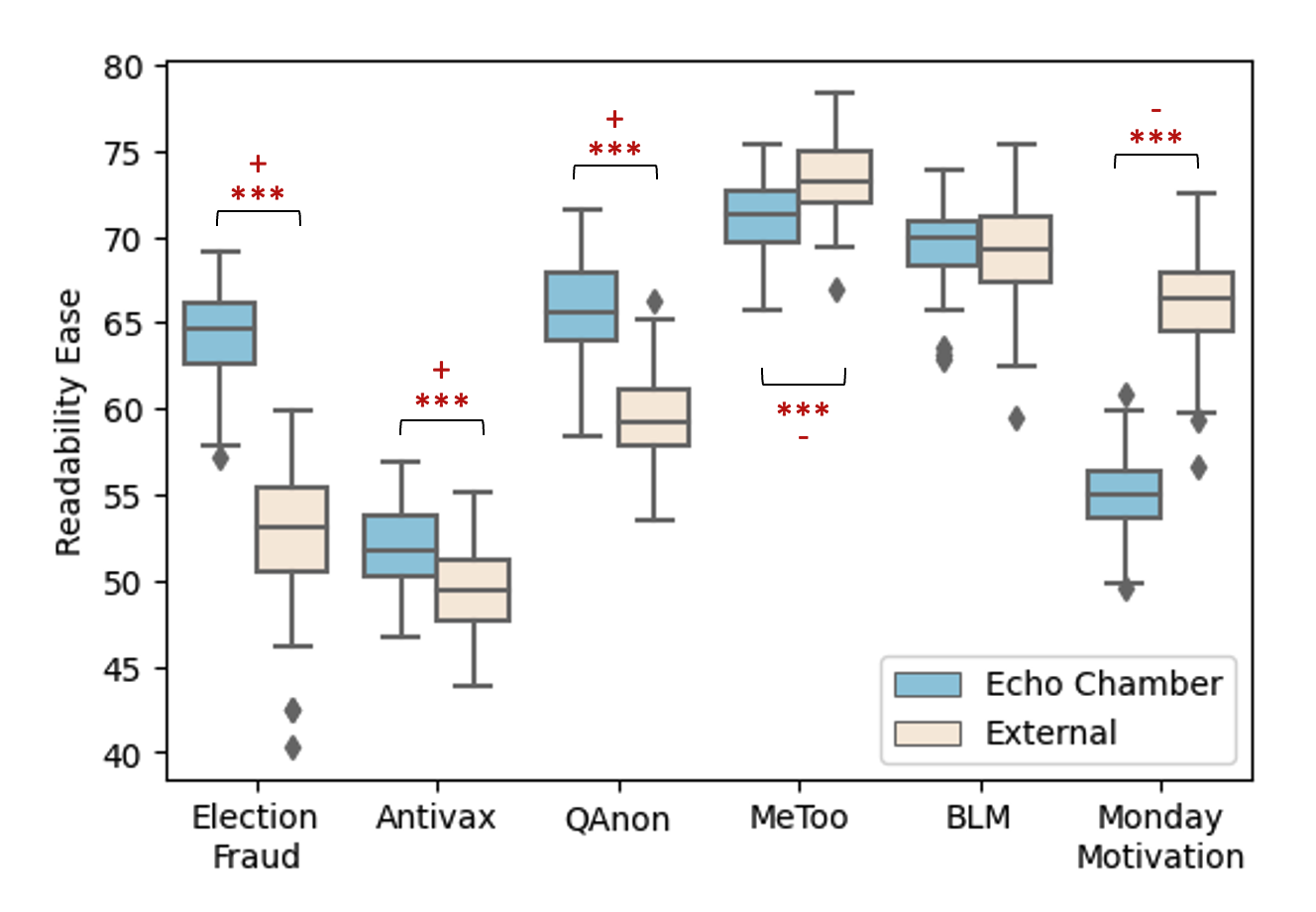}
    \caption{Readability ease score box plot for 100 samples of aggregated text. ***p<0.001, +significance in the positive direction, -significance in the negative direction. }
    \label{fig:box_plot}
\end{figure}

\vspace{0.1cm}
\noindent \emph{Readability Ease.}
Outcomes are visualized in Figure~\ref{fig:box_plot}. Content within echo chambers consistently demonstrates higher readability ease across all three misinformation topics when compared to external content. Conversely, in the cases of Metoo and Monday Motivation, a contrasting pattern emerges, indicating a higher level of complexity within echo chambers. No significant differences were observed in discussions surrounding BLM. 
Here, we emphasize the contrast between echo chambers and external members rather than making broad comparisons of readability ease between misinformation and control topics. The apparent lower overall readability ease of misinformation topics is a reflection of the topics' inherent characteristics, e.g., the complexity of these discussions or the use of specialized vocabulary.

%\textcolor{red}{I am afraid that some readers who are trying to interpret this Figure 4 in the wrong way where they are trying to compare between the overall readability ease of misinformation topics to control topics, and see that the overall level of readability ease for misinformation topics are lower than control topics. Indeed they are lower based on the specific characteristics of the topic, but what we are focusing on is the difference between echo chamber and external. I have highlighted this in Figure 4 by adding the significance of the differences using T test, but maybe we should also add something in the text to explain?} 

\vspace{0.1cm}
\noindent \emph{Big words.}
The results of all independent two-sample T-tests are presented in Table~\ref{tab:bigwords}. We also report the percentage of iterations having mean EC smaller than mean EXT and the percentage of p-value$<$0.05 in the table. Our findings consistently indicate that all misinformation topics yield a lower average percentage of big words in echo chamber posts compared to non-echo chamber posts, with all p-values less than 0.05. This suggests that echo chambers exhibit less linguistic complexity, possibly due to the development of a shared knowledge base within their community.

\begin{table*}[!ht]
    \centering
    \begin{tabular}{|c|c|c|c|c|c||c|c|}
    \hline
        \textbf{Category} & \textbf{Topic} &\textbf{Avg.EC} & \textbf{Avg. EXT}&\textbf{\makecell{T-score}}&\textbf{\makecell{p-value}}&\textbf{\makecell{\% Avg.EC\\$<$Avg.EXT}}&\textbf{\makecell{\%p-value\\$<$0.05}}\\ \hline
        \multirow{3}{*}{Misinformation} & \textbf{Election Fraud} &\textbf{22.973}&\textbf{26.638}&\textbf{-5.440}&\textbf{5.993e-8***}&\textbf{100\%}&\textbf{99.8\%}\\ \cline{2-8}
        &\textbf{Antivax }&\textbf{25.895}&\textbf{26.638}&\textbf{-2.630}& \textbf{ 0.009*}&\textbf{99.9\%}&\textbf{75.0\%}\\ \cline{2-8}
        ~ & \textbf{QAnon}&\textbf{22.044}&\textbf{23.731}&\textbf{-2.692}&\textbf{0.007*}&\textbf{99.9\%}&\textbf{ 75.0\%}\\ \hline\hline
        \multirow{2}{*}{Hyper-polarized}&MeToo &19.520&19.043&0.854&0.351&19.6\%& 12.4\%\\ \cline{2-8}
        ~&BLM &18.612&19.482&-1.156&0.120&95.6\%& 34.6\%\\ \cline{1-8}
        \makecell{Neutral} & \textbf{\makecell{Monday\\ Motivation} }&\textbf{24.133}&\textbf{19.767}&\textbf{7.239}&\textbf{6.412e-13***}&\textbf{0.0\%}&\textbf{94.7\%}\\ \hline
    \end{tabular}
    \normalsize
    \caption{Independent two-sample T-test median statistics for big words after 1000 iterations of bootstrap sampling. A significant p-value, coupled with a positive T score, indicates that echo chambers exhibit a greater utilization of big words.}
    
    \label{tab:bigwords}
\end{table*}

\subsection{Processing fluency through elaboration}

\subsubsection{Methodology}

Elaboration is captured through the use of discourse connectives.  Similar to our analysis of in-/out-group language, we employ a zero-inflated beta regression model with bootstrap sampling to examine the relationship between echo chamber membership and the use of discourse connectives.

\subsubsection{Findings}

Median statistics are shown in Table~\ref{tab:ZIBR_E}. We observe that US election fraud discourse shows a positive relationship between echo chamber membership and additive and negative connectives. QAnon exhibits a positive significant coefficient for logical and additive connectives. Conversely, in the context of the control topics Monday Motivation and BLM, an entirely opposite behavior is observed; echo chambers tend to exhibit a lower prevalence of discourse connectives, aligning with our findings above that suggest greater simplicity in these communities.

Generally speaking, we might expect fewer discourse connectives in echo chambers due to greater preexisting consensus amongst members. However, the nature of some misinformation-related topics requires more elaboration and explanation, and might explain the increased use of connectives in these contexts.

\begin{table*}[!ht]
    \centering
    
    \begin{tabular}{|c|c|c|c|c|c|c|}
    \hline
        \textbf{Category} & \textbf{Topic} &\textbf{Linguistic metrics} &\textbf{Coefficient} &\textbf{Std. Error}& \textbf{\makecell{T-score}}&\textbf{\makecell{p-value}}\\ \hline
        \multirow{9}{*}{Misinformation} & \multirow{3}{*}{Election Fraud} & Logical&0.066&0.046&1.435&0.149 \\ \cline{3-7}
        ~ & ~ & \textbf{Additive} &\textbf{0.171}&\textbf{0.043}&\textbf{4.040}&\textbf{5.600e-5***}	\\ \cline{3-7}
        ~ & ~ &\textbf{Negative}&\textbf{0.198}&\textbf{0.072}&\textbf{2.721}&\textbf{0.007*} \\ \cline{2-7}
        ~ & \multirow{3}{*}{Antivax} & Logical &-0.043&0.032&-1.340&0.180  \\ \cline{3-7}
        ~ & ~ & Additive&0.014&0.033&0.411&0.513  \\ \cline{3-7}
        ~ & ~ & Negative&-0.046&0.040&-1.148&0.248  \\ \cline{2-7}
        ~ & \multirow{3}{*}{QAnon} & \textbf{Logical} &\textbf{0.083}&  \textbf{0.042}&\textbf{1.999}&\textbf{0.046*} \\ \cline{3-7}
        ~&~&\textbf{Additive}&\textbf{0.106}&\textbf{0.040}&\textbf{2.646}&\textbf{0.008*}\\\cline{3-7}
        ~ & ~ &Negative&0.094& 0.066&1.427&0.154 \\ \hline\hline
        \multirow{6}{*}{\makecell{Hyper-polarized}} & \multirow{3}{*}{MeToo} & Logical&-0.021&0.035 &-0.592&0.453\\ \cline{3-7}
        ~&~&Additive&0.044&0.048&0.915&0.340\\\cline{3-7}
        
        ~ & ~ &Negative	&0.015&0.048 &0.305&0.512 \\ \cline{2-7}
        ~ & \multirow{3}{*}{BLM} & \textbf{Logical}&\textbf{-0.126}&\textbf{0.041}&\textbf{-3.060}&\textbf{0.002**}\\ \cline{3-7}
        ~&~&\textbf{Additive}&\textbf{-0.085}&\textbf{0.038}&\textbf{-2.264}&\textbf{0.024*}\\\cline{3-7}
        ~ & ~ &\textbf{Negative}	&\textbf{-0.118}&\textbf{0.059}&\textbf{-1.997}&\textbf{0.046*}\\ \cline{1-7}
        
        \multirow{3}{*}{\makecell{Neutral}} & \multirow{3}{*}{Monday Motivation} & \textbf{Logical}&\textbf{-0.219}&\textbf{0.040}&\textbf{-5.420}&\textbf{0.000***} \\ \cline{3-7}
        
        ~ & ~ & \textbf{Additive} &\textbf{-0.104}& \textbf{0.035}&\textbf{-2.960}&\textbf{0.003**}\\  \cline{3-7}
        ~ & ~ &\textbf{Negative}&\textbf{-0.211}	& \textbf{0.066}&\textbf{-3.182}&\textbf{0.001**}\\ \hline
       
    \end{tabular}
    
    \caption{Pairwise zero-inflated beta model median statistics of discourse connectives after bootstrap sampling(positive coefficient represents a positive association between membership in echo chambers and increased use of discourse connectives).}
    \label{tab:ZIBR_E}
\end{table*}

In line with prior work on QAnon \cite{marwick2022constructing}, \textbf{we also note cues of explanation-oriented behavior within QAnon echo chambers}, co-occuring with straightforward vocabulary. QAnon proponents often use phrases like ``do the research." These phrases appear to steer users toward an archive of QAnon-related information, likely intended for persuasive or reinforcing purposes. For instance:
\begin{quote}
\textit{There are many sides to any issue. It doesn\'t matter what others think. What matters is what you think. Do the research. Dig. Come to your own conclusion. Seek the truth knowing that you will never know what TRUTH really is, but come as close as you possibly can. \#QAnon https://t.co/W68boAmWW3}

\textit{People no longer have the luxury of claiming ignorance is why they believe the things they believe. We are living in a time where the truth can be easily accessed via documentation, Youtube, QAnon and even if you don't want to do the research their are many who do and give.}
\end{quote}

A pattern in many such tweets is the inclusion of URLs, guiding users to external sources, e.g., news articles, videos, or other tweets to ``validate'' claims. While the strategic use of archival information may vary by topic, it complements the linguistic characteristics we have identified and can be an asset for debunking misinformation.

\subsection{Case Study 2: URL usage in discussions of misinformation}
\begin{figure}[ht]
    \centering
    \includegraphics[width=\linewidth]{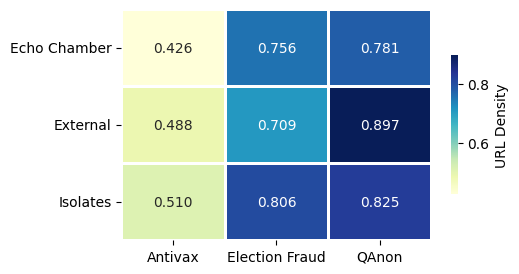}
    \caption{Heatmap of URL density of echo chamber interactions, non-echo chamber interactions (excluding isolates), and isolated tweets for misinformation topics}
    \label{fig:dist}
\end{figure}

Including URLs or referencing news sources can appear to enhance the credibility of information presented online without the need for extensive explanation.
We explore whether echo chamber members exhibit increased URL usage, as compared to external nodes. To examine this, we assess URL density in interactions within echo chambers, outside echo chambers, and in standalone original tweets across our three misinformation topics, as shown in Figure~\ref{fig:dist}. 
Notably, we observe a consistently higher URL density in isolated posts across all misinformation-related topics. Conversely, echo chamber interactions tend to utilize URLs less frequently, perhaps suggesting that users within echo chambers may not feel the need to persuade each other with new information, given their synchronized beliefs.

In between-topic comparisons, QAnon discussions have most frequent URL usage, followed by election fraud. In contrast, discussions about anti-vaccination do not engage as much with URLs. 
Among QAnon tweets, 73.68\% contains at least one URL, in comparison to Election Fraud (69.81\%) and Antivax (44.45\%).
This trend is accentuated when considering tweets with at least two URLs. Within QAnon discourse, a substantial 10.82\% of tweets fall into this category, as compared to 3.51\% for US election fraud and 3.91\% for anti-vaccination. Extensive use of URLs in QAnon discussions suggests a deliberate strategy to strengthen main assertions with multiple references.

%We also observe various cues indicative of explanation-oriented behavior. A recurring strategy among QAnon proponents is the use of phrases like "do the research." Such directives appear to guide users towards an archival base of QAnon-related information, likely intended for persuasive or reinforcing purposes. For example:
%\begin{quote}
%\textit{Do you think QAnon is a joke? That it is all crazy? I challenge you to watch this video until the end and do the research for yourself. It can't hurt to watch. https://t.co/q3rgCYzFCA}

%\textit{Open your eyes. Do your own research. Q WWG1WGA \#Q \#QANON https://t.co/aOJDMpn7F0}
%\end{quote}

\section{Discussion and Conclusion}

Our work is motivated by longstanding theories from the socio-psychologial and socio-linguistics literatures that describe phenomena like collective belonging through social identity and the importance of familiarity and fluency in information credibility. We operationalize these theories through a suite of computational metrics and analyses, which let us test these theories in the wild through social media with particular focus on discussions around misinformation-related topics.  This approach also allows us to differentially explore echo chambers and understand their linguistic underpinnings.

We do not see uniform behavior across all misinformation topics.
Perhaps surprisingly, anti-vaccination discourse shows distinct linguistic patterns from US election fraud and QAnon discussions. This may signify the differential impact of \emph{political} group identities, in particular, on the linguistic characteristics of politically-charged topics. 
We also note that the neutral topic we selected (\#mondaymotivation) consistently displays behaviors contrary to those of misinformation. This underscores that echo chamber dynamics are not uniform but are influenced by discourse.

These insights might enable identification of social communities prone to influence through anomalous linguistic signs. Moreover, understanding the anatomy of echo chambers offers a blueprint for designing more effective and targeted moderation strategies.
By studying the linguistic underpinning of social communities around misinformation-related discussions,  we offer a fresh perspective on dynamics of information dissemination in the digital age. %We open up new avenues for pinpointing vulnerable communities and enabling targeted interventions to counter misinformation. 

\section{Limitations and Future Work}
Our methodology necessitates the selection of a data collection time frame coincident with peak popularity of the respective topics. Given our intentional control over the total volume of tweets, it is an intrinsic outcome that different topics span different lengths of time. Future work should consider controlling for temporal variables to better understand the influence of time on our findings.

Additionally, here we employed a relatively strict, network structure-based definition of echo chambers using strongly connected components within topic-specific discourse. It is possible that other approaches to defining echo chambers would yield different insights. 

Finally, our study selected the zero-inflated beta model to analyze the relationship between echo chamber membership and linguistic patterns as it is suitable for the data distribution. The extensive size of our dataset results in the models being hyper-sensitive, detecting even the slightest deviations in values. This heightened sensitivity is not unique to this model but is observed in other tests we employed, such as permutation tests. As a remedy, we adopted bootstrapping across 1,000 iterations for all tests, using the median as a representative measure. Nevertheless, it is the direction and magnitude of the coefficients that offer deeper insights into the interplay between echo chamber dynamics and linguistic underpinnings in discussions.
%Finally, our comparative analyses and case studies were designed to discuss observed relationships from one potential perspective. Nonetheless, it is acknowledged that this approach may not encompass the entirety of the topic's characteristics. We concede that alternative explanations might underlie the observed phenomena and recommend further exploration in future studies.

\section{Ethical Considerations}

This study primarily involves the comprehensive analysis of data collected from social media users on a large scale. The data was obtained from the Twitter Developer API is solely utilized for the purposes of this research. We ensure that the analyses conducted do not display any identifiable information of users. Tweet IDs are shared with the public for reproducibility of the analyses in this work. Our findings are expected to support ongoing conversations within data ethics about misinformation and polarization on social media, and potential interventions. %However, we acknowledge the potential for our findings to be misused, e.g., for the strategic dissemination of misinformation online. Given that our findings are topic-focused, we consider the likelihood of intentional misuse to be low.

%\section{Acknowledgements}

\begin{acks}
This work was partially supported by NSF award \#2318460.

\end{acks}
%\section{Competing interests}
%The authors declare no competing interests. 
\bibliographystyle{ACM-Reference-Format}
\bibliography{main.bib}

\end{document}